\documentclass[10pt]{article}

\usepackage{fancybox}
\usepackage[tight,normalsize,sf,SF]{subfigure}
\usepackage[pdftex]{graphicx}
\usepackage{multirow}
\usepackage{array}
\usepackage{booktabs}
\usepackage{rotating}
\usepackage[british]{babel}
\usepackage{hyperref}

\begin{document}

\title{Discovering Dataset Nature through Algorithmic Clustering based on String Compression\thanks{This is the postprint version of an article published in IEEE TKDE. The final published version is available at \href{https://doi.org/10.1109/TKDE.2014.2345396}{https://doi.org/10.1109/TKDE.2014.2345396}. ©2015 IEEE.}}
\date{}

\maketitle

\author{Ana Granados,
        Kostadin Koroutchev,
        and Francisco de Borja Rodr\'iguez
}

\begin{abstract}
Text datasets can be represented using models that do not preserve text structure, or using models that preserve text
structure. Our hypothesis is that depending on the dataset nature, there can be advantages using a model that preserves
text structure over one that does not, and viceversa. The key is to determine the best way of representing a particular
dataset, based on the dataset itself. In this work, we propose to investigate this problem by combining text distortion
and algorithmic clustering based on string compression. Specifically, a distortion technique previously developed by
the authors is applied to destroy text structure progressively. Following this, a clustering algorithm based on string
compression is used to analyze the effects of the distortion on the information contained in the texts. Several
experiments are carried out on text datasets and artificially-generated datasets. The results show that in strongly
structural datasets the clustering results worsen as text structure is progressively destroyed. Besides, they show that
using a compressor which enables the choice of the size of the left-context symbols helps to determine the nature of
the datasets. Finally, the results are contrasted with a method based on multidimensional projections and analogous
conclusions are obtained.
\end{abstract}


\section{Introduction}
\label{Introduction}

Most text clustering methods are based on representing texts using the Vector Space Model (VSM). This model, also known as the bag-of-words model, is an algebraic model where each text document is modeled as a linear vector representing the occurrence of independent words in the text corpus \cite{Salton75}. VSM has been widely and successfully applied in many research areas. However, although representing text documents using this model makes sense from a point of view of computational efficiency, it can be imprecise because word order can be extremely important. Thus, the phrases ``the department chair couches offers'' and ``the chair department offers couches'' have the same unigram statistics, but are about quite different topics \cite{Wallach06}.

On account of that, there have been several attempts to extend the bag-of-words model. Most of these attempts are based on taking into account relationships between words. Such relationships have been dealt with using graphs \cite{Chau09}, ontologies \cite{Hotho03,Cao05}, information extraction techniques \cite{Mooney05} or topic modeling \cite{Wallach06,Wang07}. Said approaches have been applied to different research areas related to text management. Thus, among others, they have been applied to text mining \cite{Chau09,Mooney05}, text clustering \cite{Hotho03}, language modeling \cite{Wallach06,Cao05} or information retrieval \cite{Wang07}.

A natural way of taking into account relationships between words (text structure) is applying compression distances. Such distances give a measure of similarity between two objects using data compression. This means that they can give a measure of the similarity between two texts from texts themselves. In other words, texts do not need to be represented using any model, but they can be used directly. This makes text structure be considered because it is simply unvaried.

However, although text structure gives important information on a text, unmodified texts contain non-relevant words that can make the comparison of texts difficult \cite{Granados11tkde}. Due to that, we developed a distortion technique with the aim of helping the compressor obtain more reliable similarities in a compression-based clustering scenario \cite{Granados11tkde,Granados12eswa,Granados12aicomm,Granados13kais}. This distortion technique removes non-relevant information while preserving both relevant information and text structure. The way in which this is done is by removing the most frequent words in the English language from the documents, replacing each of their characters with an asterisk. This simple idea allows maintenance of text structure, while filtering the information contained in texts because, thanks to the asterisks, the lengths and the places of appearance of the removed words are maintained despite the distortion \cite{Granados12eswa}.

It has been shown that the application of that technique can improve the non-distorted clustering results \cite{Granados11tkde}. Besides, it has been shown that one of the keys to this success is that the said technique maintains text structure \cite{Granados12eswa}. In addition, it has been presented that the combination of our distortion technique with document segmentation can improve the obtained results in a compression-based retrieval scenario \cite{Granados13kais}.

In this work, we apply our distortion technique with a different purpose. In this case, we use our technique as the tool that allows the discovery of the structural characteristics of datasets, that is, the discovery of their nature. The analysis carried out to discover dataset nature can be divided into four parts. First, we study how different compression algorithms capture structure. Second, we carry out an analysis that studies how changing the size of the context affects the clustering results. Third, we analyze the dependence of the PPMD orders on the measured NCD using artificial data generated from probabilistic context-free grammars. Finally, we validate our approach by comparing it with a method based on visualizing high-dimensional data through mapping techniques. All the phases of this analysis are focused on evaluating if our approach can be used to gain an insight into the structural characteristics of datasets.

The main contributions of our work can be described as follows.

Firstly, we analyze how several compression algorithms capture text structure. In particular, LZMA, BZIP2, and PPMD are analyzed because they belong to different families of compressors: LZMA is a dictionary compressor, BZIP2 is a block-sorting compressor and PPMD is an adaptive statistical compressor. Analyzing the experimental results one can observe that destroying text structure affects the clustering behavior in a different manner depending on the dataset used. Our hypothesis is that depending on the dataset nature, there can be datasets where ignoring text structure is preferable to considering it, and vice versa. In other words, the key would be discovering if representing a particular dataset using a model which does not preserve text structure is preferable to representing it using a model that preserves text structure.

Secondly, we carry out an empirical analysis to investigate our hypothesis. This analysis is based on using a compression algorithm that allows its order to be changed. In general, an order-$N$ adaptive context-based modeler considers the $N$ symbols preceding the symbol being processed. In our analysis, PPMD compression algorithm is used to carry out the clustering because this compressor gives us the liberty to choose the number $N$ of symbols that would be taken into account to compress a symbol. Using PPMD allows us to choose the amount of text structure that will be taken into account to compute the compression distance because PPMD allows us to choose its order. The experimental results obtained in this empirical analysis show that using a bigger order is preferable when working with strongly structural datasets, whereas using a smaller order is preferable when working with datasets where the similarities between documents are captured thanks to keywords instead of text structure. In fact, they show that one can apply several PPMD orders to gain an insight into the structural characteristics of datasets, or, in other words, to gain an insight into their nature.

Thirdly, in order to better analyze this observation, we perform an analysis based on artificial data that studies the connection between the structural characteristics of datasets and the PPMD order. We have created the datasets used in this analysis by using probabilistic context-free grammars. This allows us to control the structural characteristics of datasets because the artificially-generated data has a known context given by the grammar. The fact that the context of the analyzed texts is known allows exploring if there is a correlation between the context of the texts and the PPMD order used to compute the compression distance. The experimental results show that the best results are obtained when the context of the grammar coincides with the PPMD order.

Finally, in order to corroborate that the NCD-based clustering can be used to discover the structural characteristics of datasets, we analyze how destroying text structure affects the NCDs using a method based on visualizing high-dimensional data through mapping techniques. The obtained results show that both methods behave similarly, therefore, our approach based on NCD-driven clustering is validated.

\section{Compression algorithms}
\label{Compression algorithms}

Many compression strategies have been used since the emergence of data compression as a research field, from primitive algorithms to sophisticated algorithms that achieve very high compression rates. In this work, data compression is used as the tool that allows us to give a measure of similarity between two documents. This can be done thanks to the existence of compression distances, which measure the similarity of two objects using compression algorithms, as described below.

A natural measure of similarity assumes that two objects $x$ and $y$ are similar if the basic blocks of $x$ are in $y$ and vice versa. If this happens we can describe object $x$ by making reference to the blocks belonging to $y$, thus the description of $x$ will be very simple using the description of $y$.

This is what a compressor does to code the concatenated $xy$ sequence: a search for information shared by both sequences in order to reduce the redundancy of the whole sequence. If the result is small, it means that the information contained in $x$ can be used to code $y$, following the similarity conditions described previously.

This was studied by \cite{Cilibrasi05,Li04}, giving rise to the concept of \emph{Normalized Compression Distance} (NCD), whose mathematical formulation is as follows:

\begin{eqnarray}
NCD(x,y)=\frac{\max\{C(xy)-C(x),C(yx)-C(y)\}}{\max\{C(x),C(y)\}}~,\nonumber
\end{eqnarray}

\noindent where $C$ is a compression algorithm, $C(x)$ is the size of the $C$-compressed version of $x$, $C(xy)$ is the compressed size of the concatenation of $x$ and $y$, and so on. In practice, the NCD is a non-negative number $0 \leq r \leq 1 + \varepsilon$ representing how different two objects are. Smaller numbers represent more similar objects. The $\varepsilon$ in the upper bound is due to imperfections in compression techniques, but for most standard compression algorithms, one is unlikely to see an $\varepsilon$ above 0.1 \cite{Cilibrasi05}.

The theoretical foundations for this measure can be traced back to the notion of Kolmogorov complexity $K(X)$ of a string $X$, which is the size of the shortest program able to output $X$ in a universal Turing machine \cite{Turing36,Kolmogorov65,li97}. As this function is incomputable due to the Halting problem  \cite{Sipser06}, the most usual estimation is based on data compression: $C(X)$ is considered a good upper estimate of $K(X)$, assuming that $C$ is a reasonably good compressor for $X$ \cite{Cilibrasi05}.

The NCD has been applied to several research areas because of its parameter-free nature, wide applicability and leading efficacy. For example, among others, it has been applied to data mining \cite{cilibrasi07}, clustering \cite{cilibrasi2004acm}, image analysis \cite{Cohen09}, earth observation \cite{Cerra13}, phishing detection \cite{Chen10}, or spyware detection \cite{Lavesson12}. Besides, an alternative reformulation of this measure has been used for visual analysis of document collections through mapping techniques \cite{Telles07}.

In this work we apply NCD to text clustering. In order to analyze how different compression algorithms capture text structure, three compression algorithms are used in this work to perform the NCD-driven text clustering. Each of them belongs to a different family of compressors: LZMA, PPMD, and BZIP2.

\subsection{Dictionary methods: LZMA}

Dictionary compressors break the text into fragments that are saved in a data structure called \emph{dictionary}. When a fragment of new text is found to be identical to one of the dictionary entries, a pointer to that entry is written on the compressed stream.

The most famous dictionary compressors are the ones that belong to the Lempel-Ziv family \cite{Salomon10}. The origin of this family of compressors is the LZ77 and the LZ78, which were developed by Jacob Ziv and Abraham Lempel \cite{Ziv77,Ziv78}. The LZ77 algorithm uses as part of the previously seen input stream a dictionary. The method is based on a sliding window that the encoder shifts as the strings of symbols are being encoded. The window is divided into two parts, the first part, called the \emph{search buffer}, is the current dictionary, while the second part, called the \emph{look-ahead buffer} contains the text yet to be encoded. It is important to point out that practical implementations of this method use really long \emph{search buffers} of thousands of bytes in length, and small \emph{look-ahead buffers} of tens of bytes in length \cite {Salomon10}.

In this work, the Lempel-Ziv-Markov chain algorithm LZMA \cite{lzmax}, created by Igor Pavlov, is used. This is a compression algorithm that uses a variant of the LZ77 to encode the input, and then uses a range encoder to encode the output obtained by the LZ77.

The LZMA produces a stream of literal symbols and phrase references, which is encoded one bit at a time by the range encoder, using a model to make a probability prediction of each bit. This gives much better compression because it avoids mixing unrelated bits together in the same context. In fact, empirical evidence shows that it performs very well on structured data and it looks very much like any other LZ algorithm. However, it trounces them all \cite{lzma-bloom}.

\subsection{Block-based methods: BZIP2}

BZIP2 is a block-sorting compressor developed by Julian Seward \cite{bzip2} that compresses data using Run Length Encoding (RLE), the Burrows-Wheeler Transform (BWT), the Move-To-Front (MTF) transform and Huffman coding.

The algorithm reads the input stream block by block and each block is compressed separately as one string. The length of the blocks is between 100 and 900 KB. The compressor uses the BWT to convert frequently-recurring character sequences into strings of identical letters, and then it applies MTF transform and Huffman coding. The most characteristic phase of this algorithm is the BWT, which is an algorithm created by Michael Burrows and David Wheeler \cite{Burrows94}. BWT permutes the order of the characters of the string being transformed with the purpose of bringing repetitions of the characters closer. This is useful for compression, since there are techniques such as MTF and RLE that work very well when the input string contains runs of repeated characters \cite{Salomon10}.

\subsection{Statistical methods: PPMD}

Statistical compressors are based on developing statistical models of the text. The model assigns probabilities to the input symbols, and then, the symbols are coded based on these probabilities. The model can be \emph{static} or \emph{dynamic}, depending on whether the probabilities are fixed or dynamic, that is, updated as more data is being input. The latter are more suitable because they adapt to the particularities of the data contained in the file being compressed \cite{Salomon10}.

One of the most famous statistical compressors is the PPM algorithm, whose name stands for \emph{Prediction with Partial string Matching}. PPM was originally developed by John Clearly and Ian Witten \cite{Cleary84}, with extensions and an implementation by Alistair Moffat \cite{Moffat90}. PPM is a finite-context statistical modeling technique that can be viewed as blending together several fixed-order context models to predict the next character in the input sequence. Prediction probabilities for each context in the model are calculated from frequency counts which are updated adaptively; and the symbol that actually occurs is encoded relative to its predicted distribution using arithmetic coding. The maximum context length is a fixed constant, and it has been found that increasing it beyond about six or so does not generally improve compression \cite{Cleary95}.

PPM uses sophisticated data structures and it usually achieves the best performance of any real compressor although it is also usually the slowest and most memory-intensive \cite{Cilibrasi05}. Many variants of the PPM algorithm have been implemented \cite{Salomon10}: PPMA, PPMB, PPMP, PPMX, PPMZ, etcetera. In this work, the variant called PPMD, by Dmitri Shkarin, is used because it allows us to choose the size of the context \cite{ppmd}. In general, an order-$N$ adaptive context-based modeler considers the $N$ symbols preceding the symbol being processed. In the case of the PPMD compressor that we use, the number $N$ of symbols that would be taken into account to compress a symbol can be chosen. If no number $N$ is given, $N$ takes the default value of 6.

\section{Experimental setup}
\label{Experimental setup}

This section describes firstly the distortion techniques explored in this work. Secondly it gives the details of the compression-based clustering algorithm evaluated in this work. It is important to highlight that, in this work, clustering is used as the metric that allows us to analyze how the analyzed distortion techniques destroy the structure of texts. Finally, the last subsection outlines the datasets used.

\subsection{Text processing: distortion techniques}
\label{Text processing}

In previous work, we have shown that a possible way of highlighting relevant information from texts is by removing the most frequent words in the English language from the texts, replacing each character of the removed words with an asterisk \cite{Granados11tkde,Granados12eswa,Granados12aicomm,Granados13kais}. It is important to point out that, thanks to the asterisks, the lengths and the places of appearance of the removed words are maintained despite the distortion. Hence, the application of such a distortion technique implies not only the removal of words but also the maintenance of the previous text structure. Thus, by applying such a distortion technique, both the relevant information and the structure of texts are highlighted.

In this work, we explore several distortion techniques that modify text structure from less to more. These techniques consist of first applying the distortion technique described above, and then randomly permuting different parts of the distorted texts. In order to increase the readability of the manuscript, we have given a name to each explored distortion technique. The distortion technique presented above is called \emph{Original Order (OO)} distortion technique because no random permuting is carried out after replacing the words with asterisks. The names of the other distortion techniques reflect their nature as shown below:

\begin{itemize}
\item
\emph{Randomly Permuting Asterisks (RPA)}: after replacing the words using asterisks, the strings of asterisks are randomly permuted. That is, the remaining words are maintained in their original places of appearance, while the removed words are not. It is important to note that each string of asterisks is treated as a whole. That is, if a word such as ``hello'' is replaced by ``*****'' these asterisks always remain together. This method is created to evaluate the importance of the structure of the replaced words.

\item
\emph{Randomly Permuting Remaining Words (RPRW)}: after replacing the words using asterisks, the remaining words are randomly permuted. That is, the structure of the asterisks is maintained, while the structure of the remaining words is not. This method is created to evaluate the importance of the structure of the remaining words.

\item
\emph{Randomly Permuting Everything (RPE)}: after replacing the words using asterisks, both the strings of asterisks and the remaining words are randomly permuted. It should be pointed out that, in this case, the strings of asterisks are randomly permuted as a whole too. This method, which is the most similar to the bag-of-words model, is created as a control experiment.
\end{itemize}

\subsection{Text processing: degree of distortion}
\label{Text processing: degree of distortion}

In this work, an external and well-known corpus, the British National Corpus (BNC), is used to select the words that will be removed from the documents. The BNC is a 100-million-word collection of samples of written and spoken language from a wide range of sources, designed to represent a wide cross-section of current British English, both spoken and written \cite{BNC}.

The frequencies of the English words are estimated using the BNC, and then the list of words is sorted in decreasing order of frequency. In order to study the clustering behavior evolution as the amount of removed words increases, ten sets of words are created, each one containing the words that accumulate a specific
frequency of words, these values going from 0.1 to 1.0. It is worth mentioning that each set contains the words that belong to the previous set. For example, the first set only contains the words \emph{the}, \emph{of} and \emph{and}, because these words are frequent enough to accumulate a frequency of 0.1. The second set
contains these words, together with the words necessary to accumulate a total frequency of 0.2. It has to be remarked that even when all the words included in the BNC are replaced from the texts, the words that are not included in the BNC remain in the documents. Thus, the number of words remaining in the documents when all the words included in the BNC are removed from them is 7502 for the books dataset, 119 for the UCI-KDD dataset, 98 for the Medline dataset and 96 for the IMDB dataset. It has to be pointed out that in the case of the books dataset this number is higher than in the others due to the fact that the size of the books is much bigger than the size of the texts belonging to the other datasets.

\subsection{NCD-based clustering algorithm}
\label{NCD-based clustering algorithm}

In this work, the NCD-based clustering is used as the metric that allows us to analyze how the analyzed distortion techniques destroy the structure of texts. In terms of implementation, the CompLearn Toolkit \cite{complearn} has been used to carry out the experiments. This tool implements the clustering algorithm described in \cite{Cilibrasi05,Li04} which is based on the Normalized Compression Distance (NCD). Furthermore, in order to confirm that the said NCD-based clustering can be used to discover the structural characteristics of datasets, section \ref{Visual insight} evaluates the effects that destroying text structure has on the NCDs, using a method based on visualizing high-dimensional data through mapping techniques \cite{Paulovich07}.

In the CompLearn Toolkit, the output of the clustering algorithm is represented as a dendrogram \cite{Cilibrasi05}, which is an undirected binary tree diagram that illustrates the arrangement of the clusters produced by a clustering algorithm. The way in which we quantitatively measure the error of a dendrogram is based on adding the pairwise distances between the nodes that should be clustered together. The distance between two leaves is defined as the minimum number of internal nodes needed to go from one to the other.

The procedure carried out to measure the error of a dendrogram is as follows. First, the pairwise distances between the documents that should be clustered together are added. Second, after calculating this addition, the addition that corresponds to errorless clustering is subtracted from the total quantity obtained in the first step. Consequently, the clustering error that corresponds to a dendrogram that clusters all the nodes correctly (errorless dendrogram) is 0, and the bigger the distances between the nodes that should be clustered together, the bigger the clustering error would be.

\begin{figure}[ht]
\begin{tabular}{c}
\includegraphics[width=6cm]{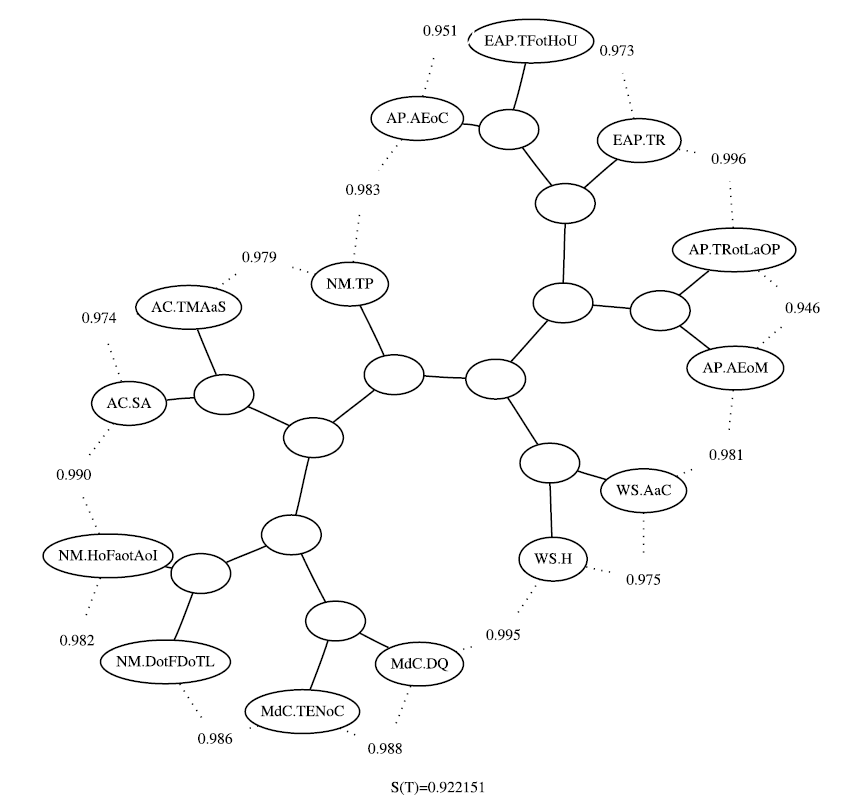} \\
  \scriptsize{(a)} \\
\includegraphics[width=6cm]{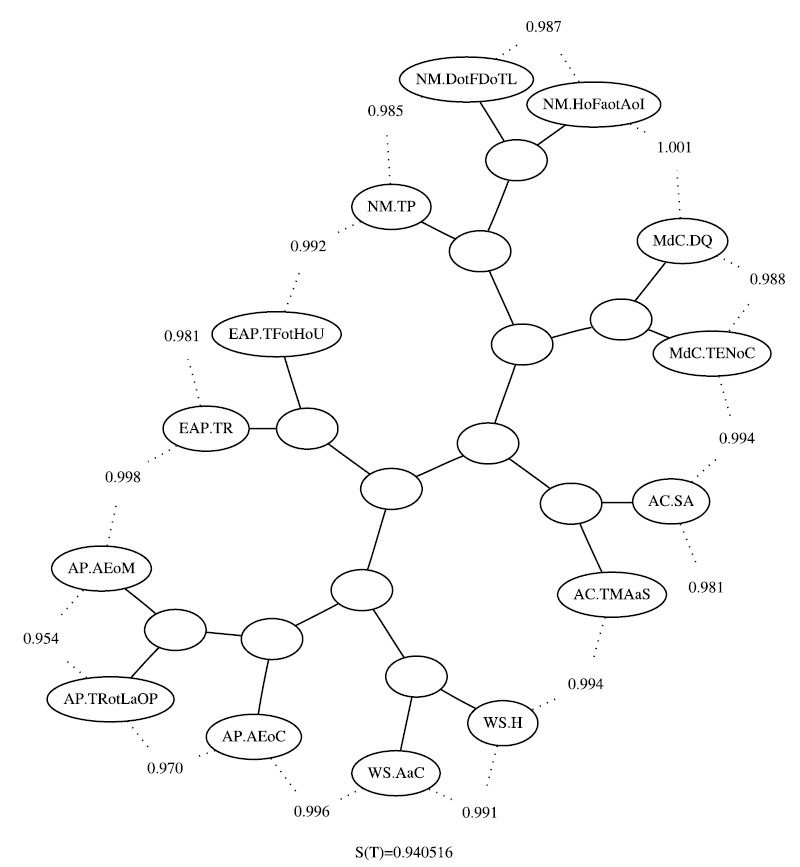} \\
   \scriptsize{(b)} \\
\end{tabular}
  \centering
\caption{\footnotesize{Example of a dendrogram for the Books repository (a). Errorless dendrogram for the Books repository (b). Each leaf of the dendrogram corresponds to a document. The nodes labeled as ``NM.TP'' and ``AP.AEoC'' are incorrectly clustered in (a). This implies that the distances between the books by Niccol\`o Machiavelli (NM) and by Alexander Pope (AP) are higher than they should be if these nodes had been correctly clustered. The pairwise distances between the nodes belonging to this dendrogram can be seen in Table \ref{Table.Dendro}.}}
\label{Fig:dendros}
\end{figure}

Fig. \ref{Fig:dendros}(a) shows an example of a dendrogram with errors, Fig. \ref{Fig:dendros}(b) shows an example of a dendrogram without errors, and Table \ref{Table.Dendro} shows the distances between all the leaves belonging to the dendrograms depicted in Fig. \ref{Fig:dendros}.

\begin{table}[ht]
\centering \caption{\footnotesize{Clustering error measurement. Pairwise distances.}} \label{Table.Dendro}
\footnotesize{
\begin{tabular}{|c|c|c|c|}
  \hline
  Cluster & Nodes & Fig.\ref{Fig:dendros}(a) \hspace{-0.3cm} & Fig.\ref{Fig:dendros}(b) \hspace{-0.3cm} \\
  \hline
  AC & AC.SA - AC.TMAaS & 1 & 1 \\
  \hline
     & AP.AEoC - AP.AEoM     & 4  & 2 \\
  AP & AP.AEoC - AP.TRotLaOP & 4  & 2 \\
     & AP.AEoM - AP.TRotLaOP & 1  & 1 \\
  \hline
  EAP & EAP.TFotHou - EAP.TR & 2  & 1 \\
  \hline
  MdC & MdC.DQ - MdC.TENoC & 1  & 1 \\
  \hline
      & NM.TP - NM.DotFDoTL        & 4  & 2 \\
  NM  & NM.TP - NM.HoFaotAoI       & 4  & 2 \\
      & NM.DotFDoTL - NM.HoFaotAoI & 1  & 1 \\
  \hline
  WS & WS.H - WS.AaC & 1  & 1 \\
\midrule
    \multicolumn{2}{|l|}{Clustering error} & 9 & 0 \\
\midrule
    \multicolumn{2}{|l|}{Dendrogram Silhouette Coefficient} & 0.589 & 0.767 \\
\midrule
\end{tabular}}
\end{table}

In addition, in order to use a well-known clustering metric apart from our clustering error measure, we calculate the Silhouette coefficient \cite{Rousseeuw87} from a dendrogram adapting its general formula to our specific domain. Before explaining the adaptation carried out, the general formula of the Silhouette coefficient has to be presented.

The Silhouette coefficient gives a measure of the quality of the clusters obtained in a clustering scenario. For each datum $i$, the following measures are calculated:

\begin{itemize}
  \item $a(i)$, which is the average dissimilarity of $i$ with all other data within the same cluster.
  \item $b(i)$, which is the lowest average dissimilarity of $i$ to any other cluster in which $i$ is not a member.
\end{itemize}

Then, using these measures, for each datum $i$, the following formula is calculated:

\begin{eqnarray}
s(i) = \frac{b(i) - a(i)}{max\{a(i),b(i)\}}.
\end{eqnarray}

It has to be remarked that a small value of $a(i)$ means that datum $i$ is similar to the rest of data within its cluster. Thus, the smaller the value of $a(i)$, the better the assignment of datum $i$. Furthermore, a large $b(i)$ means that datum $i$ is dissimilar to the data within other clusters. Thus, the greater the value of $b(i)$, the better the assignment of datum $i$. Combining these two facts one can conclude that the greater $s(i)$, the better the assignment of datum $i$.

Finally, the Silhouette coefficient of the whole dataset is calculated averaging $s(i)$ over all data of the entire dataset. The greater the Silhouette coefficient, the better the dataset has been clustered.

The Silhouette coefficient is usually calculated using cosine or Euclidean distances. However, as mentioned previously, in this work, we have adapted it to our application domain, which is based on dendrograms. This has been done by considering the distance between two datum (dendrogram leaves) as the minimum number of internal nodes needed to go from one to the other. Thus, we have defined the Dendrogram Silhouette Coefficient (DSC) using the following formula:

\begin{eqnarray}
DSC = \frac{1}{N}\sum\limits{s(i)}~,\nonumber
\end{eqnarray}

\noindent where $N$ is the number of documents contained in the dendrogram and $s(i)$ is defined as shown in formula (1).

In our case, the dissimilarity measures $a(i)$ and $b(i)$ have been defined as follows:

\begin{eqnarray}
~~a(i) = \frac{1}{n} \sum\limits_{j~\in~C_i}{d\{i,j\}} ~,~~ b(i) = \frac{1}{m} \sum\limits_{j~\notin~C_i}{d\{i,j\}}~,\nonumber
\end{eqnarray}

\noindent where $n$ is the number of elements that belong to the same cluster $C_i$ as document $i$, $m$ is the number of elements that do not belong to the same cluster as document $i$, and $d\{i,j\}$ is the pairwise distance between the documents $i$ and $j$. The latter measured in terms of the minimum number of internal nodes needed to go from $i$ to $j$. It has to be pointed out that to calculate $a(i)$ only the documents $j$ that belong to the same cluster as document $i$ have been considered, whereas to calculate $b(i)$ only the documents $j$ that do not belong to the same cluster as document $i$ have been considered.

The Silhouette coefficients that correspond to the dendrograms depicted in Fig. \ref{Fig:dendros}(a) and \ref{Fig:dendros}(b) are 0.589 and 0.767 respectively. One might think that this measure can be inconvenient because a dendrogram that clusters perfectly all the documents has a DSC of 0.767. In that sense, our clustering error would be better. However, DSC is a powerful tool that allows us to summarize and compare clustering results that have been obtained in different experiments and different datasets, as will be shown in section \ref{Experimental results}. In other words, each measure has its strengths, that is the reason why we use both.

\subsection{Datasets}
\label{Datasets}

The same datasets used in \cite{Granados11tkde,Granados12eswa} are used in this work. That is, four datasets composed of texts written in English have been used in the experimental phase. Their description is as follows:

\begin{itemize}
\item
Books: Fourteen classical books from universal literature, to be clustered by author. Since the documents belonging to this dataset are books, their size is
quite large in general.

\item
UCI-KDD: Sixteen messages from a newsgroup (UCI-KDD) \cite{uci-kdd}, to be clustered by topic. The main characteristic of these texts is their reduced size.

\item
Medline: Twelve documents from the Medline repository \cite{medlineplus}, to be clustered by topic. Since these texts are about medicine, they are very specific and their vocabulary is very specialized.

\item
IMDB: Fourteen plots of movies from the Internet Movie Data Base (IMDB) \cite{imdb}, to be clustered by saga. An important characteristic of these documents is the presence of names of characters and places that are related to the sagas.
\end{itemize}

\section{Experimental results}
\label{Experimental results}

This section analyzes first the effects of applying the distortion techniques presented in subsection \ref{Text processing}, using PPMD and BZIP2 compressors to calculate NCD. Then it uses PPMD to analyze more deeply the hypothesis that in datasets where the NCD captures the similarities between documents thanks to the structure of the documents, the structure of asterisks is very important. This is done by investigating how using different context sizes affect the results in a certain way, depending on the nature of the dataset. To this end, several experiments are carried out on both text datasets and artificially-generated datasets. It has to be pointed out that PPMD is used for this purpose because it allows a change in the size of the context. Finally, an evaluation on how destroying text structure affects the NCDs is carried out. For this purpose, a method based on visualizing high-dimensional data through mapping techniques is used.

\subsection{Analyzing different compression algorithms}
\label{Analyzing different compression algorithms}

In previous work \cite{Granados12eswa} we analyzed the effects of applying the distortion techniques presented in subsection \ref{Text processing}, using the LZMA compression algorithm to calculate the NCD, and therefore, using the LZMA to perform the NCD-based clustering method. In this work we extend that work by using the PPMD and the BZIP2 compressors to calculate the NCD. We perform such analysis with the purpose of analyzing how several compression algorithms capture text structure.

\begin{figure}[h!t]
\begin{tabular}{cc}
\hspace{-0.5cm} \includegraphics[angle=270,width=4.4cm]{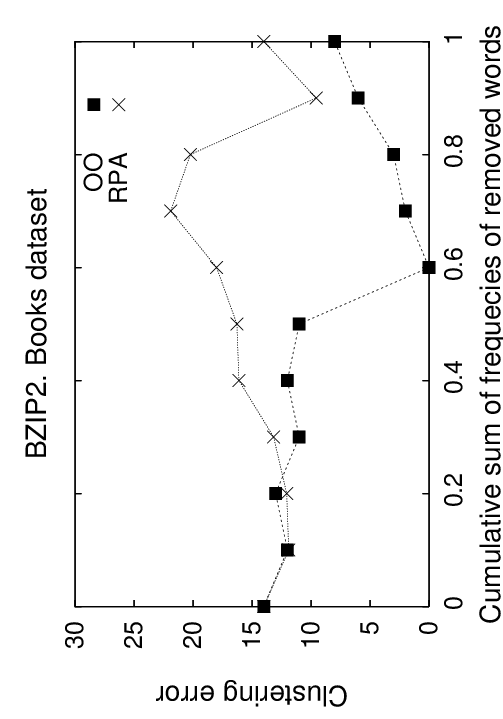} & \hspace{-0.6cm}
\includegraphics[angle=270,width=4.4cm]{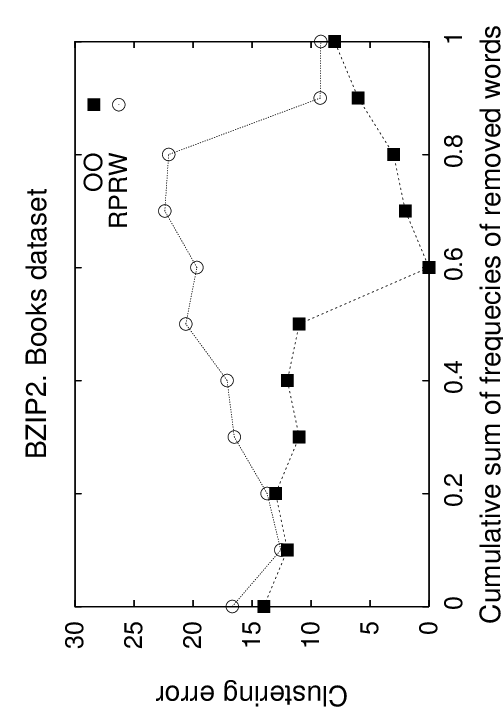} \\
\scriptsize{(a) RPA distortion technique} &  \scriptsize{(b) RPRW distortion technique} \\
\hspace{-0.5cm} \includegraphics[angle=270,width=4.4cm]{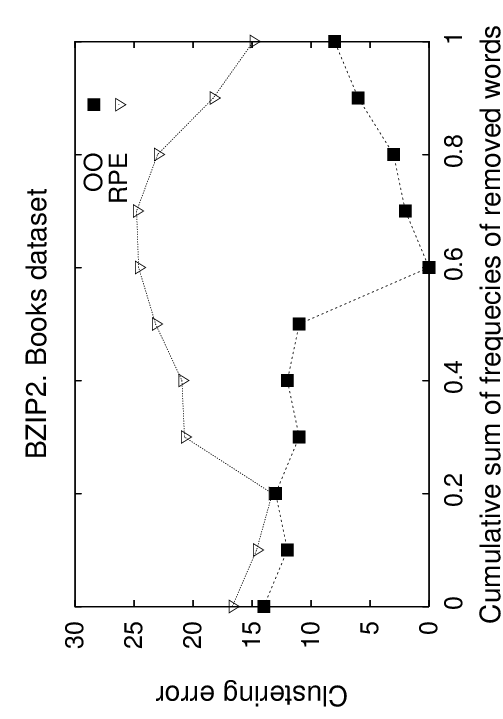} & \hspace{-0.6cm}
\includegraphics[angle=270,width=4.4cm]{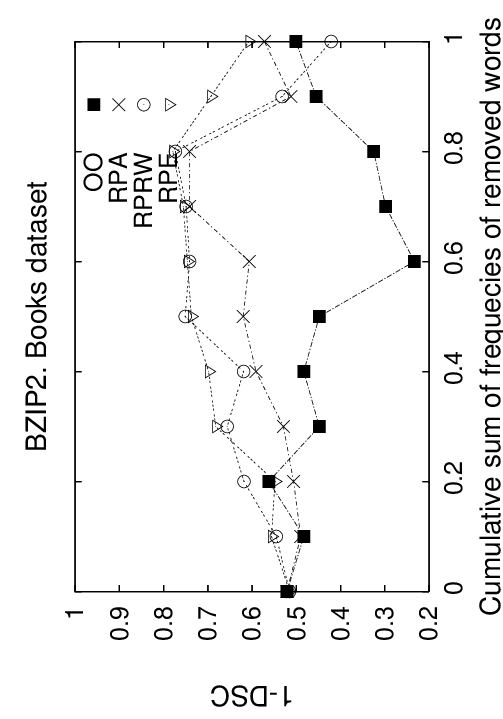}  \\
\scriptsize{(c) RPE distortion technique} & \scriptsize{(d) RPA, RPRW and RPE} \\
\end{tabular}
  \centering
\caption{\footnotesize{Clustering results for the Books dataset and the BZIP2.}}
  \label{Fig:books-bzip2}
\end{figure}

As a first sample, and to gain intuition, this section shows graphically the results obtained using both compressors on the Books dataset. A figure is shown for each compression algorithm. In each figure, the clustering error obtained applying the \emph{Original Order (OO)} distortion technique is plotted in panels (a), (b) and (c), in order to allow comparison between this technique and the rest of the distortion techniques. These panels correspond to the \emph{Randomly Permuting Asterisks (RPA)}, \emph{Randomly Permuting Remaining Words (RPRW)}, and \emph{Randomly Permuting Everything (RPE)} distortion techniques, respectively (see subsection \ref{Text processing} to further details on the distortion techniques). The mean of the clustering error is depicted in these last panels because their corresponding distortion techniques are based on permuting different parts of the texts, and therefore, the experiments have been repeated twelve times (it has to be remarked that non-significative changes have been observed for different number of repetitions). In the plots depicted in panels (a), (b) and (c), the value on the vertical axis corresponds to the obtained clustering error, whereas the value on the horizontal axis corresponds to the cumulative sum of frequencies of the words that are removed from the texts. Besides, in order to contrast our clustering error with a well-known clustering metric, a measure based on the Silhouette Coefficient \cite{Rousseeuw87} has been created and shown in panel (d) of Fig. \ref{Fig:books-bzip2}.

Thus, in order to analyze the experimental results in a more global way, two measures that summarize the clustering errors obtained for each distortion technique and each dataset are created. These measures are $DSC_i$ and $DSC^r_i$.

The DSC that summarizes the clustering quality of a distortion technique $i$ is calculated by averaging the DSC obtained for all the distortions $d$ when the distortion technique $i$ is applied:

\begin{eqnarray}
DSC_i = \frac{1}{10} \sum_{d=0.1}^{1.0}{DSC_{i,d}}~,\nonumber
\end{eqnarray}

\noindent where $i$ could be one of the distortion techniques presented in subsection \ref{Text processing}, $d$ is the cumulative sum of frequencies of the words that are removed from the texts, $d$ going from 0.1 to 1.0, as explained in section \ref{Text processing: degree of distortion}, and $DSC_{i,d}$ is the $DSC$ obtained for distortion technique $i$ and distortion $d$.

In order to ease the comparison of the qualities of the clustering results obtained using the distortion techniques that destroy text structure with the technique that maintains text structure, the following measure has been created:

\begin{eqnarray}
DSC^r_i = \frac{DSC_{OO} - DSC_i}{1 - DSC_i}~,\nonumber
\end{eqnarray}

\noindent where $DSC_{OO}$ is the quality of the clustering results obtained for the \emph{Original Order} distortion technique, and $DSC_i$ is the quality of the clustering results obtained for the distortion technique $i$. $DSC^r_i$ allows us to compare the results obtained applying the distortion technique $i$ ($DSC_i$) with the ones obtained applying the \emph{Original Order} distortion technique ($DSC_{OO}$). Therefore, $DSC^r_{OO}$ is always 0, since $DSC_i = DSC_{OO}$. Furthermore, it has to be highlighted that the greater $DSC^r_i$ the worse the results obtained for the $i$ distortion technique with respect to the \emph{Original Order} distortion technique.

\begin{figure}[h!t]
\begin{tabular}{cc}
\hspace{-0.5cm} \includegraphics[angle=270,width=4.4cm]{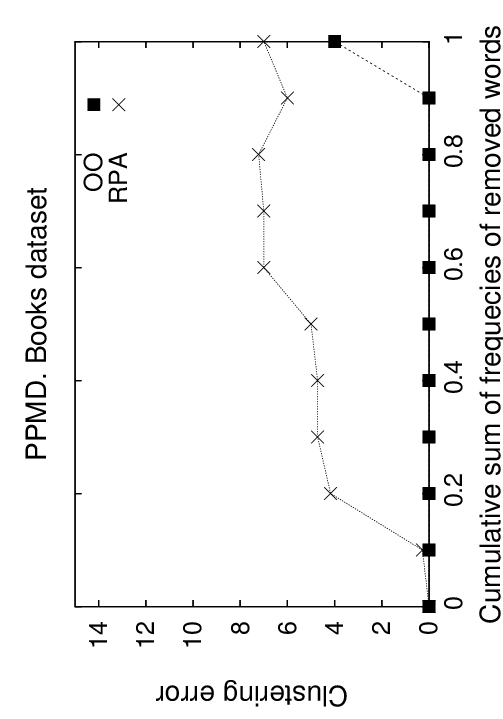} & \hspace{-0.6cm}
\includegraphics[angle=270,width=4.4cm]{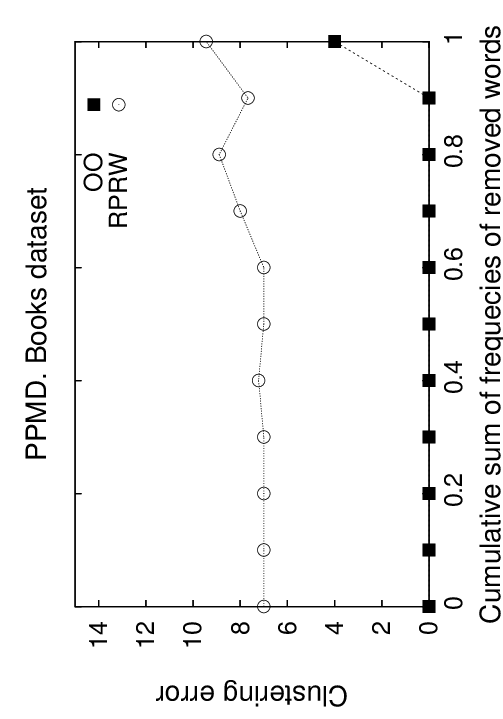} \\
\scriptsize{(a) RPA distortion technique} &  \scriptsize{(b) RPRW distortion technique} \\
\hspace{-0.5cm} \includegraphics[angle=270,width=4.4cm]{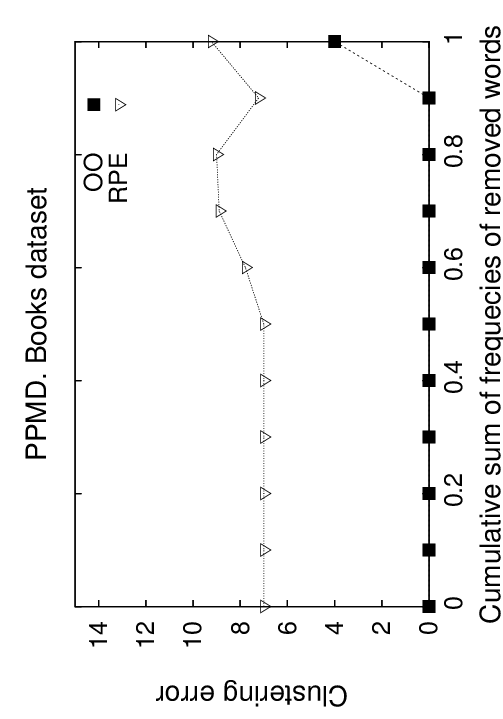} & \hspace{-0.6cm}
\includegraphics[angle=270,width=4.4cm]{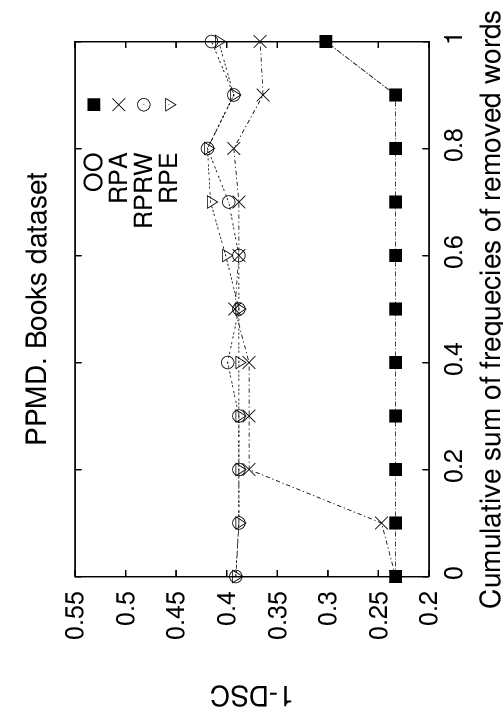}  \\
\scriptsize{(c) RPE distortion technique} & \scriptsize{(d) RPA, RPRW and RPE} \\
\end{tabular}
  \centering
\caption{\footnotesize{Clustering results for the Books dataset and the PPMD (order 6) compression algorithm. Panels (a), (b) and (c) show the clustering error obtained for all the distortion techniques. Panel (d) shows the evolution of $1-DSC$ for all the distortion techniques. The clustering error gets worse as text structure is destroyed. This behavior is observed when both clustering error measures are used.}}
  \label{Fig:books-ppmd}
\end{figure}

Fig. \ref{Fig:books-ppmd} shows the results that correspond to the PPMD compression algorithm. When the structure of texts is not lost, the clustering error remains constant from 0.0 to 0.9, as can be observed looking at the curve that corresponds to the \emph{Original Order (OO)} distortion technique. It is important to remark that, in this case, the non-distorted clustering error cannot be improved since its value is 0. Interesting conclusions can be drawn comparing the said curve with the others. First, losing the structure of texts makes the clustering results get worse as the amount of removed words increases, as panel (a) shows. Second, losing the remaining words' structure makes the clustering results get worse when the texts contain a lot of remaining words and a few of asterisks, as panel (b) shows. Third, losing every structure, both behaviors are observed at the same time, that is, the clustering errors are worse for small and large numbers of removed words, as shown in panel (c). Finally, comparing panels (a), (b) and (c) with panel (d) one can observe that our clustering error measure behaves exactly the same as $1-DSC$ (note that DSC gives a measure of the quality of the clustering, that is the reason that to obtain analogous curves $1-DSC$ has to be computed).

The same behavior is observed when the BZIP2 compression algorithm is used (see Fig. \ref{Fig:books-bzip2}). In this case, the non-distorted clustering error can be improved because it is greater than 0. In fact, it is improved when the structure of texts is maintained, reaching a clustering error of 0, as the curve that corresponds to the \emph{Original Order (OO)} distortion technique shows. Therefore, it seems that preserving the structure of texts while filtering the information contained in them, the compressor obtains more reliable similarities, and the non-distorted clustering results can be improved.

In addition, by way of example, an in order to gain intuition on how $DSC_i$ and $DSC^r_i$ behave, table \ref{Table.NCE} analyzes how $DSC^r_i$ summarizes the results shown in Fig. \ref{Fig:books-bzip2}. Comparing the clustering error obtained applying the \emph{Original Order (OO)} distortion technique with the rest of distortion techniques, one can observe that $DSC_{RPA}$ is smaller than $DSC_{OO}$, $DSC_{RPRW}$ is smaller than $DSC_{RPA}$, and $DSC_{RPE}$ is smaller than $DSC_{RPRW}$. This is reflected in the corresponding $DSC^r_i$ values.

\begin{table}[h]
\centering \caption{\footnotesize{$DSC_i$ and $DSC^r_i$ for the results shown in Fig. \ref{Fig:books-bzip2}.} 
\label{Table.NCE}}
\footnotesize{
\begin{tabular}{|l|l|}
  \hline
  \textbf{$DSC_i$}      & \textbf{$DSC^r_i$}  \\
  \hline
  $DSC_{OO}$ = 0.576    & $DSC^r_{OO}$ = 0.000   \\
  $DSC_{RPA}$ = 0.406   & $DSC^r_{RPA}$ = 0.287  \\
  $DSC_{RPRW}$ = 0.364  & $DSC^r_{RPRW}$ = 0.334 \\
  $DSC_{RPE}$ = 0.321   & $DSC^r_{RPE}$ = 0.376  \\
  \hline
\end{tabular}}
\end{table}

\begin{figure*}[t]
\begin{tabular}{ccc}
\hspace{-2.5cm} \includegraphics[width=5.6cm]{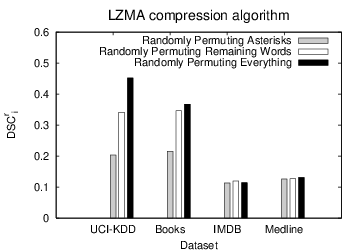} &
\includegraphics[width=5.6cm]{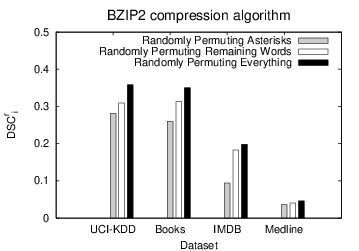} &
\includegraphics[width=5.6cm]{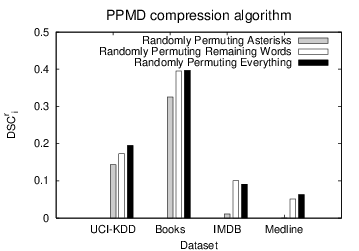} \\
\end{tabular}
  \centering
\caption{\footnotesize{Visual representation of the structural differences between the datasets. $DSC^r_{i}$ values for all the datasets and all the compressors using the distortion techniques RPA, RPRW and RPE. It can be observed that Books dataset and UCI-KDD dataset behave similarly, whereas IMDB dataset behaves like Medline dataset.}}
  \label{Fig:resumen-figuras-barras}
\end{figure*}

The measure $DSC^r_i$ has been used to average the results obtained for all the degrees of distortion for the distortion technique $i$. Thus, this measure summarizes the experimental results obtained for every dataset, every compression algorithm, and every distortion technique. All the values of $DSC^r_i$ are presented in Table \ref{Table.Summary}. Besides, in order to give a more visual representation of the structural differences between the datasets, these values are depicted in Fig. \ref{Fig:resumen-figuras-barras} as well.

\begin{table}[h]
\centering \caption{\footnotesize{Summary of results for all the datasets and all compression algorithms using the distortion techniques RPA, RPRW and RPE.}} 
\label{Table.Summary}
\footnotesize{
\begin{tabular}{|c|c|c|c|c|}
 \cline {2-5}
  \multicolumn{1}{c|}{$DSC^r_i$}  & \multirow{2}{*}{\textbf{Dist. Tech.}} & \multirow{2}{*}{\textbf{LZMA}} & \multirow{2}{*}{\textbf{BZIP2}} &  \textbf{PPMD} \\
  \multicolumn{1}{c|}{} & & & & \textbf{(order 6)} \\
 \hline
                 &   RPA          & 0.203 & 0.281 & 0.143 \\
\textbf{UCI-KDD} &   RPRW         & 0.341 & 0.309 & 0.173 \\
                 &   RPE          & 0.452 & 0.358 & 0.195 \\
 \hline
                 &   RPA          & 0.215 & 0.260 & 0.325 \\
\textbf{Books}   &   RPRW         & 0.346 & 0.313 & 0.395 \\
                 &   RPE          & 0.367 & 0.350 & 0.397 \\
 \hline
                 &   RPA          & 0.113 & 0.094 & 0.011 \\
\textbf{IMDB}    &   RPRW         & 0.119 & 0.183 & 0.100 \\
                 &   RPE          & 0.114 & 0.198 & 0.091 \\
 \hline
                 &   RPA          & 0.126 & 0.036 & -0.010 \\
\textbf{Medline} &   RPRW     & 0.127 & 0.040 & 0.051 \\
                 &   RPE          & 0.131 & 0.046 & 0.063 \\
 \hline
\end{tabular}}
\end{table}

Analyzing the values contained in Table \ref{Table.Summary} and the values depicted in Fig. \ref{Fig:resumen-figuras-barras}, one can observe that losing text structure affects the clustering behavior especially when the Books and the UCI-KDD datasets are used. Our hypothesis is that this occurs because the structure of the texts that comprise these datasets is highly representative to cluster them. On the contrary, losing text structure does not affect the clustering behavior so strongly in the Medline and the IMDB datasets. Our hypothesis is that this is due to the fact that the remaining words are the key factor to cluster the documents belonging to these datasets.

Summarizing, our hypothesis is that in datasets where the NCD-based clustering method works thanks to the fact that the NCD captures the structure of similar documents, text structure is very important. These are the Books and the UCI-KDD datasets. On the contrary, in datasets where the NCD captures the similarities between documents thanks to the keywords, the text structure is not so relevant. These are the Medline and the IMDB datasets. Of course, these are just two hypotheses that need to be analyzed more deeply. The next subsection analyzes precisely that.

\subsection{Analyzing different context sizes}
\label{Analyzing different context sizes}

The PPMD compression algorithm is used with the purpose of analyzing the hypothesis presented in subsection \ref{Analyzing different compression algorithms} more deeply. We use this compression algorithm because it allows a change in the size of the context, that is, the order of the modeler. In general, an order-$N$ adaptive context-based modeler considers the $N$ symbols preceding the symbol being processed. The PPMD compressor allows us to choose the number $N$ of symbols that would be taken into account to develop the statistical model of the text. If no number $N$ is given, $N$ takes the default value of 6.

Different $N$ values are used in this experimental phase with the aim of exploring the consequences of changing the order in the clustering error. These values go from 2 to 6. The maximum value of $N$ is 6 because it has been found that increasing the maximum context length beyond about six does not generally improve compression \cite{Cleary95}.

\begin{figure}[h!t]
\begin{tabular}{cc}
\hspace{-0.5cm} \includegraphics[angle=270,width=4.4cm]{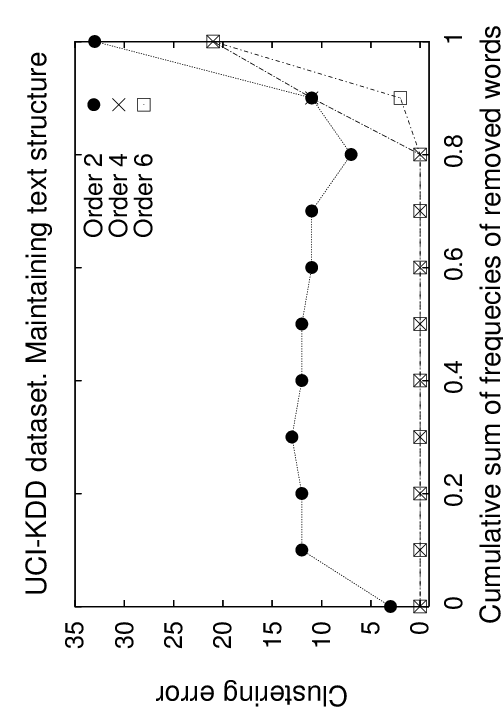} & \hspace{-0.6cm}
\includegraphics[angle=270,width=4.4cm]{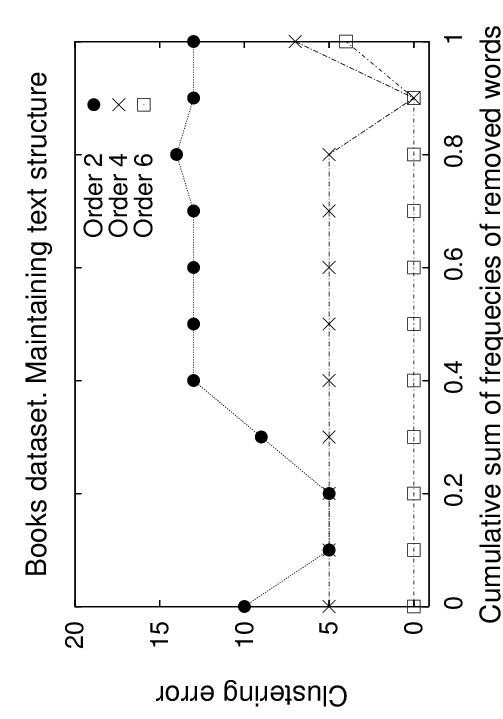} \\
\hspace{-0.5cm} \includegraphics[angle=270,width=4.4cm]{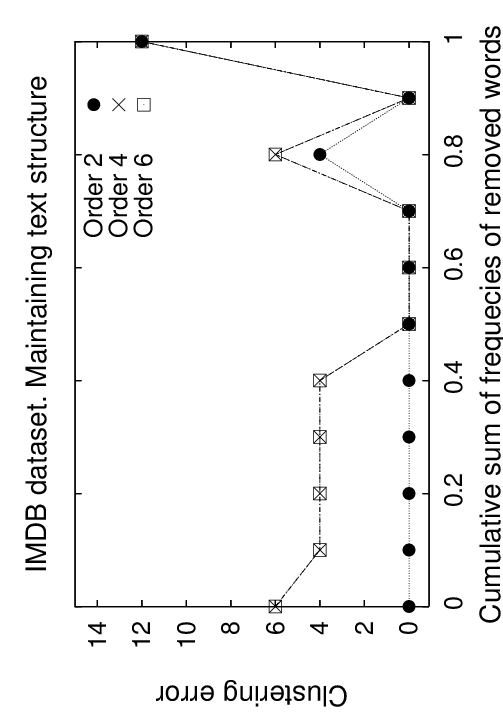} & \hspace{-0.6cm}
\includegraphics[angle=270,width=4.4cm]{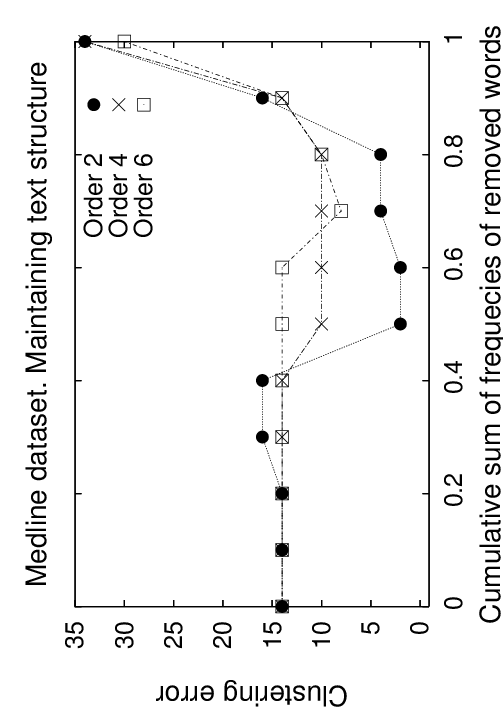}  \\
\end{tabular}
  \centering
\caption{\footnotesize{PPMD order analysis. One can observe that depending on the nature of the dataset, the best clustering results are obtained using a PPMD order of 6 (UCI-KDD and Books) or a PPMD order of 2 (IMDB and Medline).}}
  \label{Fig:ppmdorder-analysis}
\end{figure}

The hypothesis presented in subsection \ref{Analyzing different compression algorithms} is that in datasets, such as Books and UCI-KDD, where the NCD captures the similarities between documents thanks to the document structures, the contextual information is very important. This means that using very small contexts, the clustering results should be worse than using bigger contexts, because almost no contextual information is taken into account using small contexts. Our hypothesis is supported by the results depicted in Fig. \ref{Fig:ppmdorder-analysis}. Analyzing this figure one can observe that using small contexts makes the clustering error worse than using the default context size.

The contrary happens in datasets where the NCD captures the similarities between documents thanks mainly to the keywords, as we predicted in subsection \ref{Analyzing different compression algorithms}. Thus, the results depicted in Fig. \ref{Fig:ppmdorder-analysis} show that the best clustering error is obtained using small contexts. In fact, using a context of 2 when clustering the documents belonging to the Medline dataset, the clustering error is much better than using the default context size.

\begin{figure}[h!t]
\begin{tabular}{cc}
\hspace{-0.5cm} \includegraphics[angle=0,width=4.4cm]{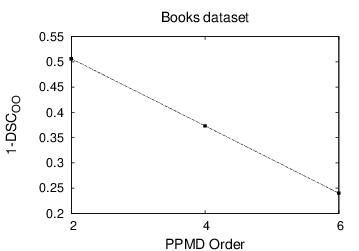} & \hspace{-0.6cm}
\includegraphics[angle=0,width=4.4cm]{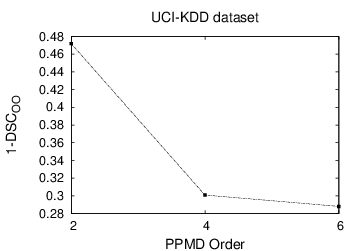} \\
\hspace{-0.5cm} \includegraphics[angle=0,width=4.4cm]{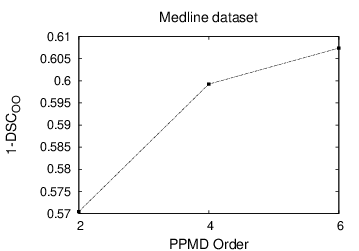} & \hspace{-0.6cm}
\includegraphics[angle=0,width=4.4cm]{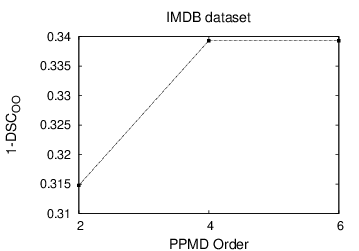}  \\
\end{tabular}
  \centering
\caption{\footnotesize{PPMD order analysis. These curves show how the quantity $1-DSC_{OO}$ changes when different PPMD orders are used. It can be observed that for datasets with the same nature, the behaviour of $1-DSC_{OO}$ is qualitatively similar.}}
  \label{Fig:NCD-summary}
\end{figure}

The evolution of the clustering quality ($DSC_i$) with respect to the PPMD order is shown in Fig. \ref{Fig:NCD-summary}. Analyzing the curves obtained for all the datasets one can observe that the form of the curve depends on the nature of the dataset. Thus, in the datasets where the NCD captures the similarities between documents thanks to the documents structure (Books and UCI-KDD) the best quality is obtained using a context of 6, and the quality decreases as the context used decreases. On the contrary, in datasets where the NCD captures the similarities between documents thanks mainly to the keywords (Medline and IMDB) the best quality is obtained at a context of 2. In this kind of dataset, where document structure is not so relevant, compressor such as LZMA or BZIP2 might be used instead of PPMD because the former are faster than the latter.
It has to be highlighted that the bigger context used is 6 because it has been found that increasing the maximum context length beyond about 6 does not generally improve compression \cite{Cleary95}.

\subsection{Artificial data}
\label{Artificial data}

Analyzing the experimental results shown in previous subsections, one can observe that the behavior observed for the Books and the UCI-KDD datasets is different from the behavior observed for the Medline and the IMDB datasets. Our hypothesis is that the behavior depends on the nature of the dataset. Thus, using a bigger context is preferable when working with strongly structural datasets, and vice versa. In this section, we carry out a control experiment to take a step towards better analyzing our hypothesis.

Our control experiment is based on the use of artificially-generated data. These kinds of data allow us to control the context because the artificial texts are generated using a context-free grammar of a known context \cite{Booth73,Manning99}. In other words, we know the context length of the explored data because we have generated them. An example of a context-free grammar is given in Table \ref{Table.grammar-real}. Using this definition, the grammar production $p\quad X: Y Z$ is equivalent to $P(Y Z|X)=p$. As usual, the language symbols that are present only in the right parts of the productions are terminals and the ones that are present also in the left parts of the productions are non-terminals.

The generation starts from the root symbol and proceeds with the probability in the first column. By construction, the generated sentences have fixed length and the entropy for each generated sentence can be computed just as $-\log_2 P$, where $P$ is the multiplication of the probabilities of the branches chosen to generate that sequence. For example, if $v=1/2$, using the grammar shown in Table \ref{Table.grammar-real}, the sentence \emph{acft.123456} is generated applying the following rules:

\begin{tabular}{ll}
  1      & SS : S .123456 \\
  v(0.5) & S : a A0  \\
  0.5    & A0 : c D2  \\
  0.5    & D2 : f D1  \\
  0.5    & D1 : t  \\
\end{tabular}

This means that the entropy of sentence \emph{acft.123456} would be $-\log_2~\frac{1}{16}$ bits because the multiplication of the probabilities of all the branches chosen to generate the sentence is $\frac{1}{16}$.

It has to be remarked that we use the delimiter .123456 to end the sentence. If the delimiter were shorter or missing, for example just '.', a problem would arise: as a result of the overlap of the boundaries of the sentences, the compressor will try to interpret totally independent probability distribution on both sides of the delimiters of the sentences. This will introduce significant error in the estimation of the entropy. In order to eliminate this sentence-boundary effect, we have added long delimiters between the sentences. Adding these delimiters, statistically, the clustering should be correct.

\begin{table}[ht]
\centering \caption{\footnotesize{The context-3 grammar used in the simulation with $v={1/4, 1/5, 1/6}, w=1/2$.}} \label{Table.grammar-real}
\footnotesize{
\begin{tabular}{clcclcclccl}
\hline

\hspace{-0.3cm} 1 & SS  : S .123456
\\
\hspace{-0.3cm} v & S : a A0 &&
0.5 & D1 : t &&
0.5 & H2 : o H0\hspace{-0.3cm}
\\
\hspace{-0.3cm} 1-v & S : b B0 &&
0.5 & D1 : u &&
0.5 & H2 : p H1\hspace{-0.3cm}
\\
\hspace{-0.3cm} 0.5 & A0 : c D2 &&
0.5 & E0 : w &&
0.5 & G0 : A
\\
\hspace{-0.3cm} 0.5 & A0 : d E2 &&
0.5 & E0 : x &&
0.5 & G0 : B
\\
\hspace{-0.3cm} 0.5 & D2 : e D0 &&
0.5 & E1 : z &&
0.5 & G1 : C
\\
\hspace{-0.3cm} 0.5 & D2 : f D1 &&
0.5 & E1 : @ &&
0.5 & G1 : D
\\
\hspace{-0.3cm} 0.5 & E2 : g E0 &&
0.5 & B0 : k G2 &&
0.5 & H0 : E
\\
\hspace{-0.3cm} 0.5 & E2 : h E1 &&
0.5 & B0 : l H2 &&
0.5 & H0 : F
\\
\hspace{-0.3cm} w & D0 : r &&
0.5 & G2 : m G0 &&
0.5 & H1 : G
\\
\hspace{-0.3cm} 1-w & D0 : s &&
0.5 & G2 : n G1 &&
0.5 & H1 : H
\\
\hline
\end{tabular}}
\end{table}

In the grammar shown in Table \ref{Table.grammar-real}, we use as a delimiter the string $.123456$ (in the left part of the root non-terminal production SS), which is longer than the PPMD's order. Using such grammars we can generate texts with context-length defined by the grammar. For example, the language generated from the grammar given in Table \ref{Table.grammar-real} would have a left context-length of 3 by construction, because the maximum depth of the grammar tree in this case is 4 and all production of SS has length 4 plus the length of the delimiter. Indeed we can predict any next symbol using the three preceding symbols with the probability given in the first column of the language definition.

\begin{figure*}[t]
\begin{tabular}{ccc}
\hspace{-2.9cm}\includegraphics[width=4.2cm]{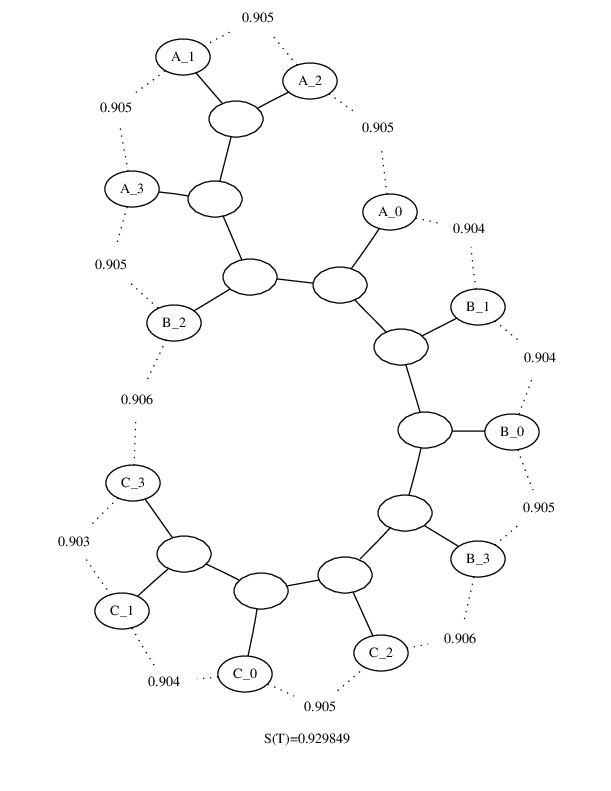} & \includegraphics[width=6cm]{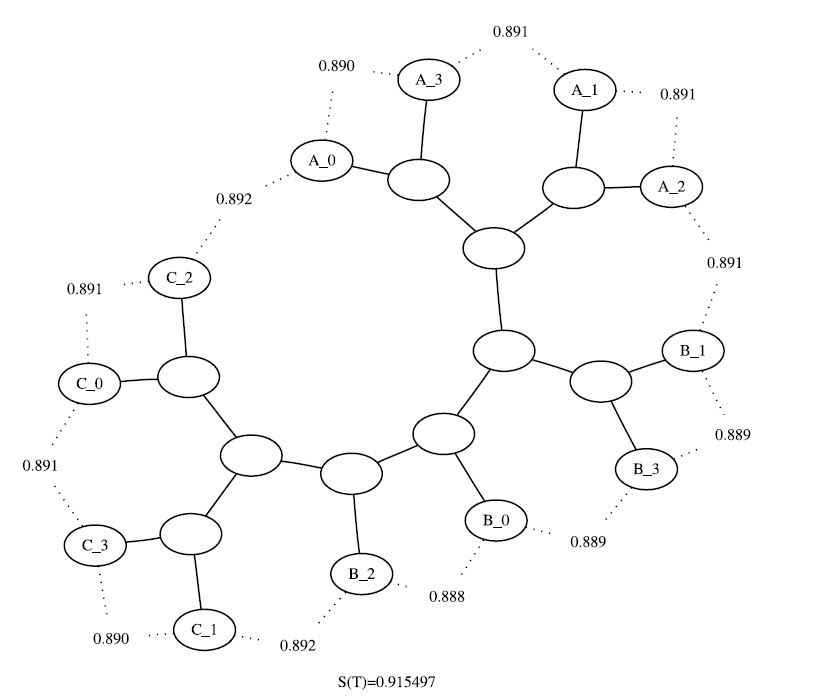} & \includegraphics[width=6cm]{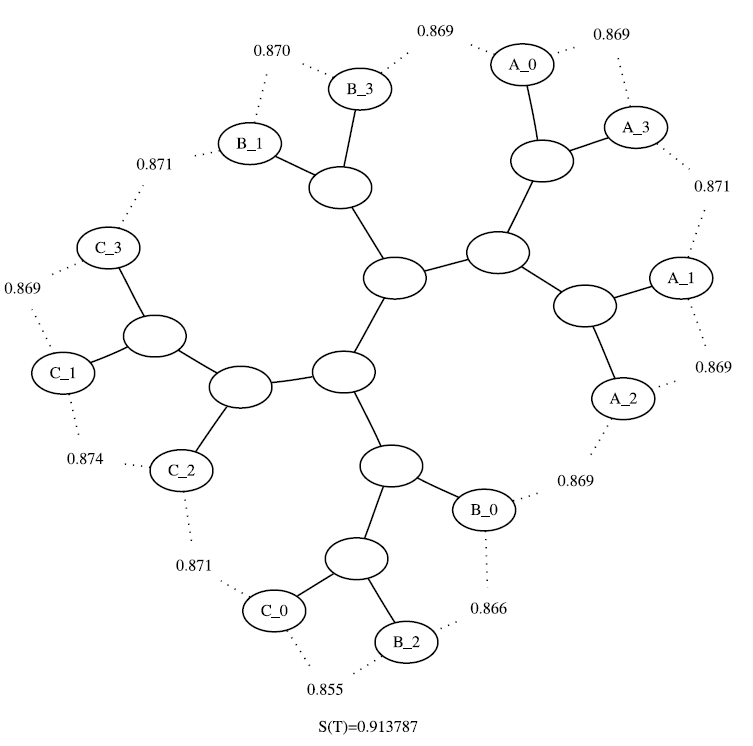} \\
 \scriptsize{(a) PPMD order = 2} & \scriptsize{(c) PPMD order = 4} & \scriptsize{(c) PPMD order = 6} \\
\end{tabular}
  \centering
\caption{\footnotesize{Context of language = 4. All the nodes are correctly clustered when the context of the grammar coincides with the PPMD order. }}
  \label{Fig:dendros-ox}
\end{figure*}

We can calculate the probability of each of the sentences generated by the context-free probabilistic grammars \cite{Stolcke94}. Consequently, we can calculate the expectation of the entropy of a sufficiently long text. Given that the grammars we used in this article are designed so that one sentence can only have a unique grammar decomposition, the calculation is trivial. Namely, the probability is always the product of the probabilities of the grammatical productions. Furthermore, if we present a sufficient number of sentences of the grammar to the PPMD compressor (with a delimiter longer than the context allowed by PPMD), we can have quite a precise estimation of the conditional probabilities derived by the grammar.

If a grammar $G_A$ provides probabilities $\{p_X\}$ for the sentences $X$ and the grammar $G_B$ provides probabilities $\{q_X\}$, extracted from the samples $A$ and $B$ respectively, we can say that if we compress the concatenation $AB$ then, at least in the beginning of $B$, we will compress the set $q$ with the set $p$, so we can expect a mean length of the sentence $X$ to be $-q_X\log_2 p_X$ bits instead of the optimal $-q_X\log_2 q_X$ bits. This will produce a difference between the optimal length and the one achieved in the compression equal to $q_X\log_2 q_X/p_X$. This is exactly the Kullback distance \cite{Kullback51} between the probability distribution $\{p\}$ and $\{q\}$. Therefore, we can have a theoretical measure of what can be expected as the difference. The estimation can be done more precisely, although it is beyond the scope of this article. However, we can state that if two probabilistic languages have different conditional probabilities, the more the probabilities differ, the more the length the compressed concatenation will increase with respect to the optimal coding, e.g. the NCD will be larger.

In order to confirm this hypothesis, we generate texts according to the probabilistic grammar given in table \ref{Table.grammar-real}, assigning different values of the parameters $v$ and $w$. For each set of values ($v$,$w$) we generate sentences and concatenate them in files of size 16000 bytes. This procedure gives uniformity to the artificial data that cannot be achieved using human-generated texts. Depending on the probabilities given to the rules of the grammar, the texts generated belong to one or another cluster. The probabilities used to generate the texts have been selected to make elements belonging to different clusters very similar. In fact, the difference between them is less than the difference between difficult-to-distinguish human dialects.

As an example, Fig. \ref{Fig:dendros-ox} shows the dendrograms obtained when clustering artificial texts generated from a grammar of context 4, using the following probabilities to create the texts which belong to the three clusters A, B and C: $v = \frac{1}{4}$, $\frac{1}{5}$ and $\frac{1}{6}$. In all figures the first letter represents the cluster (A, B or C). Note that the differences between A, B and C cases are just 2\% to 5\% of the terminal occurrences, which is under the usual fluctuations for human-generated texts. We intentionally use such close values of the parameters, because otherwise the clustering with compression has very small error and the samples will be classified with virtually no error.

Analyzing the figures, one can observe that the classification is errorless with context 4, whereas it is not with
context of 2 and 6. That is, all the texts are correctly clustered when the context of the grammar coincides with the
PPMD order, whereas this does not occur using other contexts. The same can be observed by projection methods, as
mentioned in the next section, where we have the highest the Silhouette coefficient for the PPMD order that coincides
with the context of the grammar. It is noteworthy that projection methods allow us to represent more samples in
visually resilient way.

Summarizing, if the PPMD order is equal to the context of the artificial language, we can expect PPMD to construct an exact model of the language. If we choose a PPMD order smaller than the context of the language, the clustering will not be optimal. Finally, if we choose a PPMD order larger than the context of the language and a short delimiter, we will merge consecutive sentences of the language in the PPMD's model. This will also result in an increase in the imprecision of the NCD. All these effects are observed in the experimental results obtained.

It has to be remarked that this is an illustrative section that only shows some representative examples with the aim of gaining intuition about the relation between the context of the language and the PPMD order. In this sense, three dendrograms are shown (see Fig. \ref{Fig:dendros-ox}), each of them presenting the results of clustering twelve documents, using each time a different PPMD order to compute the NCD distance matrix. We have decided to show dendrograms of such size because we want them to be legible. However, a complete analysis of the whole set of artificially-generated texts is presented in the next subsection.

\subsection{Visual insight}
\label{Visual insight}

In previous sections we have used the NCD-based clustering as a tool that allows us to discover the structural characteristics of datasets. Roughly speaking, this has been done by measuring the quality of the obtained clustering results as text structure has been progressively destroyed. In order to corroborate that the clustering can be used for this purpose, we compare it with a different method based on visualizing high-dimensional data through mapping techniques.

Several ways of visualizing high-dimensional data through mapping techniques exist in the literature
\cite{Hinton02,Paulovich07,VanDerMaaten08}. In general terms, these mapping techniques give each datapoint a location
in a two or three dimensional map. This allows a human to visually analyze datasets because they are represented in a
two or three dimensional map, which means that the datasets are human comprehensive. In terms of implementation, we use
the Projection Explorer (PEx), a visualization tool that takes a distance matrix as input and generates a projection
through mapping techniques \cite{Paulovich07,Telles07}.

\begin{figure}[h!t]
\begin{tabular}{cc}
\includegraphics[width=4cm]{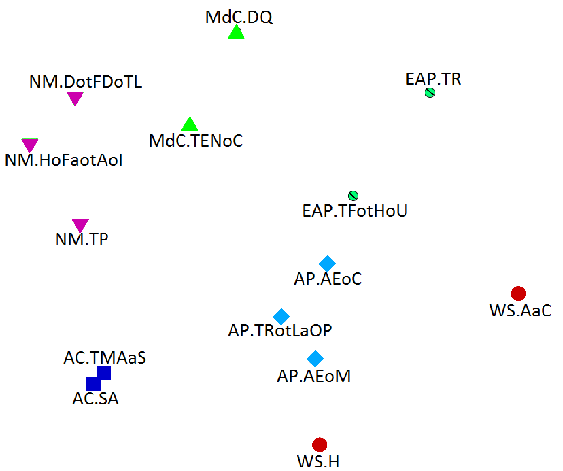} &
\includegraphics[width=4cm]{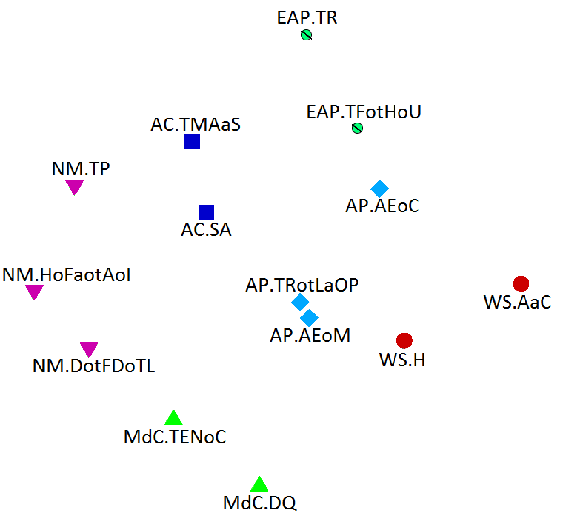} \\
\hspace{-0.4cm} \scriptsize{(a) No distortion. SC = 0.63} & \scriptsize{(b) OO. Distortion 0.9. SC = 0.61} \\
\multicolumn{2}{c}{\includegraphics[width=4cm]{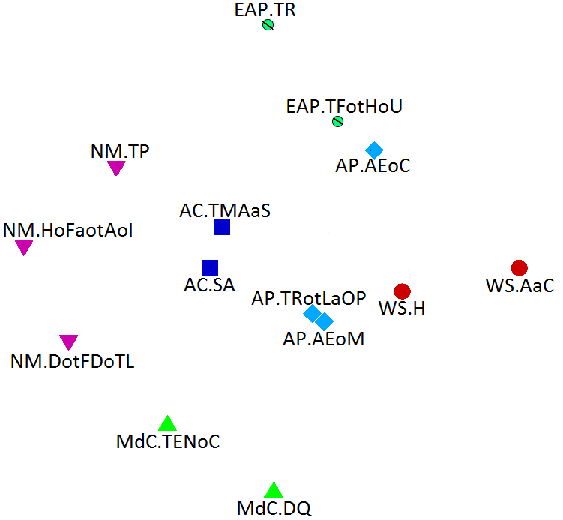}} \\
\multicolumn{2}{c}{\scriptsize{(c) RPRW. Distortion 0.8. SC = 0.55}} \\
\end{tabular}
  \centering
\caption{\footnotesize{Visual analysis of distortion. Books dataset and PPMD compressor. The worst clustering quality is obtained when the \emph{RPRW distortion technique} is applied (c). This behavior is the same as the one presented in Fig. \ref{Fig:books-ppmd}.}}
  \label{Fig.PEx.DistortionAnalysis}
\end{figure}

\begin{figure}[h!t]
\begin{tabular}{cc}
\includegraphics[width=3.5cm]{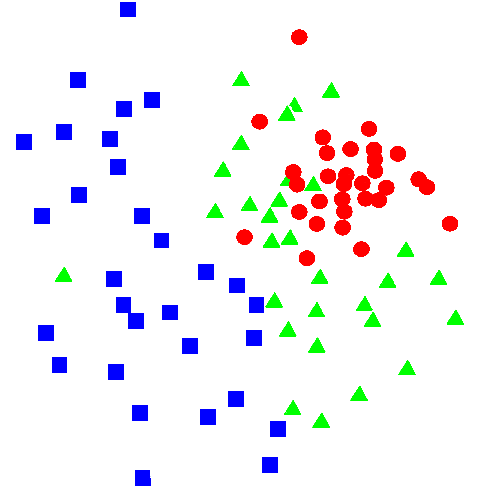} &
\includegraphics[width=3.5cm]{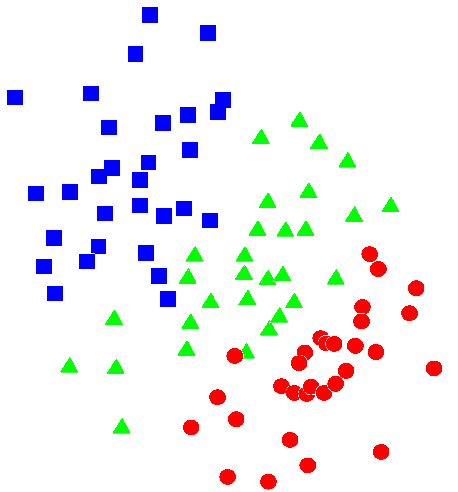} \\
\scriptsize{(a) PPMD order = 2. SC = 0.19} & \scriptsize{(b) PPMD order = 4. SC = 0.31} \\
\multicolumn{2}{c}{\includegraphics[width=3.5cm]{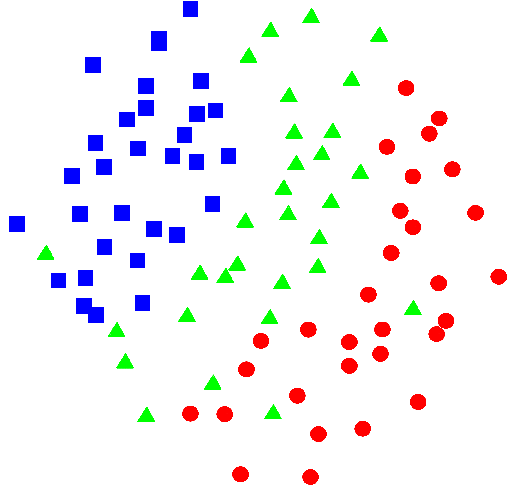}} \\
 \multicolumn{2}{c}{\scriptsize{(c) PPMD order = 6. SC = 0.23}} \\
\end{tabular}
  \centering
\caption{\footnotesize{Artificial texts with context of language = 4. Using a PPMD order that coincides with the context of the language gives the best results both using projections and clustering (see Fig. \ref{Fig:dendros-ox}).}}
  \label{Fig.PEx.ArtificialData}
\end{figure}

Firstly, we visually show the effects of distortion in the Books dataset in Fig.\ref{Fig.PEx.DistortionAnalysis}. This
figure shows the projections obtained from the distance files which correspond to the NCDs calculated from the original
books (a), the books distorted using the \emph{OO distortion technique}, for a distortion of 0.9 (b), and the books
distorted using the \emph{RPRW distortion technique}, for a distortion of 0.8 (c). Analyzing Fig.
\ref{Fig.PEx.DistortionAnalysis} one can observe that the clustering quality obtained with no-distortion and with the
\emph{OO distortion technique} for a distortion of 0.9 is similar, besides it can be observed that the worst clustering
quality is obtained when the \emph{RPRW distortion technique} is applied, as was to be expected given the results shown
in Fig. \ref{Fig:books-ppmd}.

Finally, in order to evaluate how using different PPMD orders affects the NCDs, we provide a figure
(Fig.\ref{Fig.PEx.ArtificialData}) that shows several projections made from all the artificially-generated data created
as explained in section \ref{Artificial data}. That is, these projections allow a complete evaluation of all the
artificial data generated. Thus, these projections have been made from 90 artificially-generated texts which have been
created from a context-4 grammar. A different PPMD order has been used to create the distance matrix for each
projection, these orders being 2, 4 and 6. Analyzing Fig. \ref{Fig.PEx.ArtificialData}, one can observe that the best
results are obtained for the projection that shows the NCD distances calculated using a PPMD order of 4. For the other
projections, the quality of the clustering is worse. In other words, the best results are obtained when the context of
the grammar coincides with the PPMD order, as shown for some representative examples in section \ref{Artificial data}.
Therefore, the same behavior obtained using clustering (see Fig. \ref{Fig:dendros-ox}) has been obtained using
projections (Fig. \ref{Fig.PEx.ArtificialData}).

Summarizing, the conclusions derived from our initial approach (NCD-based clustering) and this new approach
(multidimensional projections) are qualitatively independent on the method used, which makes the results presented in
the paper more reliable.

\section{Conclusions}
\label{Conclusions}

Discovering the nature of datasets can help us manage texts more efficiently. Thus, there can be datasets where
representing texts applying a model that does not preserve text structure is preferable to applying a model that
preserves it, and viceversa. In this work we suggest using a distortion technique previously developed by the authors
\cite{Granados11tkde,Granados12eswa,Granados12aicomm,Granados13kais} with the purpose of finding out what the best way
of managing a dataset is.

The idea is as follows: our distortion technique has been applied to destroy text structure progressively. After that,
the distorted texts have been clustered with the aim of analyzing the effects of the distortion on the quality of the
obtained clustering results. If clustering results get worse as text structure is destroyed, then, for this particular
dataset, applying a model that preserves text structure can be preferable. On the contrary, if clustering results are
maintained, then using a simplified model that does not preserve text structure could be suitable.

In terms of implementation, the CompLearn Toolkit \cite{complearn}, which implements the clustering algorithm described
in \cite{Cilibrasi05,Li04}, is used to perform the clustering. This clustering algorithm uses the Normalized
Compression Distance (NCD) as similarity distance between two objects. Detailed information on the NCD can be found in
section \ref{Compression algorithms}.

Since NCD has been used to measure the similarity between texts, several compression algorithms have been used to
compute the NCD. The experimental results have been found to be consistent across compression algorithms belonging to
different families of compressors. These results have shown that destroying text structure affects the clustering
behavior in a different manner depending on the dataset used. Our hypothesis is that in datasets where the structure of
documents is the key factor to cluster the documents, the contextual information is more important than in datasets
where keywords are the key factor.

An empirical analysis has been carried out with the purpose of exploring said hypothesis. The empirical analysis has
explored if changing the size of the context affects the clustering results. The PPMD compression algorithm has been
used to carry out this analysis, because this compressor allows one to choose the size of the context. The experimental
results obtained in this empirical analysis have shown that using a bigger context is preferable when working with
strongly structural datasets, whereas using a smaller context is preferable when working with datasets where the
similarities between documents are captured thanks to keywords instead of text structure. In fact, it has been shown
that one can apply several PPMD orders to gain an insight into the structural characteristics of datasets (see Fig.
\ref{Fig:NCD-summary}).

Besides, we have performed an analysis based on artificial data that studies the dependence of the PPMD orders on the
measured compression distance. These kinds of data allow us to control the context because the artificial texts are
generated using a context-free grammar of a known context. After generating the artificial texts, we have clustered
them using different PPMD orders each time. The experimental results have shown that the best results are obtained when
the context of the grammar coincides with the PPMD order. Therefore, the more similar the PPMD order and the context of
the artificial language are, the more accurate the model of the language constructed by PPMD is. These results
corroborate the experimental results obtained using real texts thanks to the fact that the context of the grammar is
known (i.e. controlled) when working with our artificial texts.

Finally, in order to corroborate that the NCD-based clustering can be used to determine the structural characteristics
of datasets, we have analyzed how destroying text structure affects the NCDs using a method based on visualizing
high-dimensional data through mapping techniques \cite{Paulovich07}. The obtained results have shown that both methods
behave similarly, therefore, our approach based on NCD-driven clustering has been validated.

Summarizing, four main contributions have been presented in the paper. Firstly, we have analyzed empirically why the
effects of applying the distortion techniques depend on the dataset used. The experimental results have shown that in
datasets where the structure of documents is the key factor to clustering the documents, the contextual information is
more important than in datasets where document structure is not so relevant. Secondly, we have carried out an analysis
that studies how changing the size of the context affects the clustering results. This analysis has been made using the
PPMD compression algorithm because this compressor gives us the liberty to choose its order (i.e. the number of symbols
that would be taken into account to compress a symbol). The experimental results have shown that we can gain an insight
into the structural characteristics of datasets by applying several PPMD orders. Thirdly, we have analyzed the
dependence of the PPMD orders on the measured NCD using artificial data generated from probabilistic context-free
grammars, finding that the best clustering results are obtained when the context of the grammar is equal to the context
taken into account by the compression algorithm. Finally, we have confirmed that our approach is correct by comparing
it with a method based on visualizing high-dimensional data through mapping techniques.

\section*{Acknowledgments}
They would like to thank the anonymous reviewers for their constructive comments on the manuscript. This work was
supported by the Spanish projects TIN2010-19607 and TIN2010-21575-C01.


\end{document}